\documentclass[apj]{emulateapj}

\begin{document}


\title{Interactions of the magnetospheres of stars and close-in giant planets}


\author{O. Cohen\altaffilmark{1}, J.J. Drake\altaffilmark{1}, V.L. Kashyap\altaffilmark{1}, 
S.H. Saar\altaffilmark{1}, I.V. Sokolov\altaffilmark{2}, W.B. Manchester IV\altaffilmark{2}, 
K.C. Hansen\altaffilmark{2}, and T.I. Gombosi\altaffilmark{2}}

\altaffiltext{1}{Harvard-Smithsonian Center for Astrophysics, 60 Garden St. Cambridge, MA 02138}
\altaffiltext{2}{Center for Space Environment Modeling, University of Michigan, 2455 Hayward St., 
Ann Arbor, MI 48109}

\begin{abstract}
Since the first discovery of an extrasolar planetary system more than a decade ago, hundreds more have 
been discovered. Surprisingly, many of these systems harbor Jupiter-class gas giants located close to 
the central star, at distances of 0.1~AU or less. Observations of chromospheric 'hot spots' that 
rotate in phase with the planetary orbit, and elevated stellar X-ray luminosities, suggest that 
these close-in planets significantly affect the structure of the outer atmosphere of the star through 
interactions between the stellar magnetic field and the planetary magnetosphere. Here we carry out the 
first detailed three-dimensional MagnetoHydroHynamics (MHD) simulation containing the two magnetic bodies and 
explore the consequences of such interactions on the steady-state coronal structure. The simulations 
reproduce the observable features of 1) increase in the total X-ray luminosity, 2) appearance of coronal hot spots, 
and 3) phase shift of these spots with respect to the direction of the planet. The proximate cause of 
these is an increase in the density of coronal plasma in the direction of the planet, which prevents 
the corona from expanding and leaking away this plasma via a stellar wind. The simulations produce 
significant low temperature heating. By including dynamical effects, such as the planetary orbital 
motion, the simulation should better reproduce the observed coronal heating.

\end{abstract}


%%%%%%%%%%%%%%%%%%%%%%%%%%%%%%%%%%%%%%%%%%%%%%%%%%%%%%%%%%%%%%%%%%%%%%%%%%%%%%
% Section - Introduction
%%%%%%%%%%%%%%%%%%%%%%%%%%%%%%%%%%%%%%%%%%%%%%%%%%%%%%%%%%%%%%%%%%%%%%%%%%%%%%

\section{INTRODUCTION}
\label{sec:Intro}

The structure and heating of the solar corona, as well as the acceleration of the 
solar wind, are influenced by the structure and topology of the large-scale coronal 
magnetic field. On this basis, the existence of a planet at a distance of 0.1~AU or 
less \citep{mayor95, exoplanets03}, with a strong internal magnetic field is expected 
to have a significant effect on the stellar magnetosphere, which is 
controlled by the magnetic field structure \citep{cuntz00}. In recent years, some signatures of this star-planet 
interaction (SPI) have been observed.  Shkolnik et al. \citep{shkolnik03, shkolnik05a,shkolnik05b,shkolnik08} 
have reported on modulations in the Ca II K emission line, an indicator for chromospheric activity. They 
find enhancements in the line intensity that have the same period as the planetary orbital 
motion, though sometimes with a significant non-zero phase-shift. The cause is deemed magnetic 
and not tidal because of the lack of an equivalent hot spot offset in phase by $180^\circ$. 
In addition, a statistical survey of the X-ray fluxes from stars with close-in planets has found them 
enhanced by a 30-400\% on average over typical fluxes from similar stars with planets that are 
not close-in \citep{kashyap08}. Direct X-ray observations of the HD 179949 system \citep{saar08} 
showed that the SPI contributed $\approx$30\% to the emission at a mean temperature of $\approx$ 1 keV.

Some analytical and semi-empirical arguments have been advanced to explain these 
observations. One posits that particles are accelerated along magnetic field lines that connect 
the star and planet, creating hot spots where they hit the chromospheric layer 
\citep{cuntz00,lanza08,cranmer07}. As a result, hot spots are observed generally in phase with 
the planetary orbit, but with the capacity to have 
large offsets, depending on the exact structure of the magnetic field between the star and planet. 
Another shows that transition of field lines from a high-helicity state to a linear force-free 
state is energetically adequate to power the enhanced intensities \citep{lanza09}. The detailed behavior of 
the dynamical interaction of coronal and wind plasma with two magnetic field systems is, however, 
very difficult to realize with idealized models. The problem properly requires simultaneous 
descriptions of both the stellar and the planetary magnetospheres, the planetary orbital motion, 
and often asynchronous stellar rotation, together with a self-consistent stellar wind solution.

Here we describe an initial simulation of the magnetic star-planet interaction. We use idealized test 
cases to study the fundamental changes in the steady-state coronal structure due to the presence of the 
planet and its magnetic field. The dynamical interaction due to the planetary orbital motion is captured 
in an indirect manner.

%%%%%%%%%%%%%%%%%%%%%%%%%%%%%%%%%%%%%%%%%%%%%%%%%%%%%%%%%%%%%%%%%%%%%%%%%%%%%%
% Section - Simulation
%%%%%%%%%%%%%%%%%%%%%%%%%%%%%%%%%%%%%%%%%%%%%%%%%%%%%%%%%%%%%%%%%%%%%%%%%%%%%%

\section{SIMULATION}
\label{sec:Simulation}

The numerical simulation has been performed using the University of Michigan Solar Corona 
(SC) model \citep{cohen07}, which is based on the BATS-R-US global MHD code \citep{powell99} and 
is part of the Space Weather Modeling Framework (SWMF) \citep{toth05}. The model solves the set 
of magnetohydrodynamic equations on a Cartesian grid using adaptive mesh refinement (AMR) technology. 
This model has been extensively validated for the solar corona using coronal observations and in-situ 
solar wind measurements taken at 1~AU \citep{cohen08}. We assume that the particular physical 
description of the coronal heating and wind acceleration is not crucial to study the change in the 
existing coronal structure due to the planet. It is important to mention that we use a {\it global} model 
for the corona that cannot reproduce realistic chromospheric emission due to heating of coronal loops. We also 
do not fully describe the observed coronal heating, since for example no input from magnetic reconnection or 
loop footpoint motion is included. Thus, while we adopt the physical parameters of some 
real systems in the modeling, we do not expect the models to fully reproduce all aspects of 
observations (in particular, details of the temperature and level of the emissions) at this point.
The full physical description of the model and its limitations can be found in \cite{cohen07,cohen08} 

We performed several different numerical simulations, of which we highlight two here. 'Case A': 
both the stellar and planetary magnetic fields are perfectly aligned dipoles. We set the 
stellar polar field to be $5G$ and the planetary polar field to be antiparallel at $-2G$ (i.e., 
opposite to the stellar dipole). The planetary magnetic field is weaker than Jupiter's, 
and follows the assumption that hot-Jupiters 
are expected (but not required) to have lower spin rates due to tidal locking, and thus have weaker magnetic fields 
\citep{sanches-lavega04,griessmeier04,olson06}. We note that a simulation in which the planetary 
dipole was set to be in the same direction with the stellar dipole resulted in a quantitatively similar 
solution as in this case. 'Case B': the planetary magnetic field is a perfect dipole and the 
stellar magnetic field is driven by solar magnetic synoptic map (magnetogram). This map contains measurements 
of the photospheric radial magnetic field taken during solar maximum (Carrington Rotation CR2010, very active Sun). 
The use of a magnetic synoptic map enables us to generate a realistic, Sun-like, three-dimensional magnetic field.

In Case A, we mimic the relative motion between the planet and the background plasma by fixing the 
planet and rotating the star and the coronal plasma in the inertial frame. This way, the planet orbits 
the star  backwards in the frame rotating with the star. This is done due to the fact that the actual 
orbital motion of the planet requires time-dependent boundary conditions. We plan to implement this 
technical improvement in future simulations. For the sake of definiteness, we partially match the 
parameters of the system to the observed parameters of HD 179949 \citep{exoplanets03}, which is an F8V type star. 
We use the following stellar parameters: $M_\star=1.28~M_\odot$, $R_\star=1.19~R_\odot$, and stellar rotation period of 
$3~d$. In the HD 179949 system, the planet is located at a distance of 0.045~AU (9.65 stellar radii), and a $60^\circ$ 
phase-lead of a chromospheric hot spot is observed. The planetary parameter $Msini=0.98~M_J$ has not been used here. In Case B, we 
fix the planet relative to the star and run the simulation in the frame rotating with the star. In this case we use solar 
parameters except for the planetary properties, which are the same as in Case A. This case represents a 
steady-state, large scale interaction of a Sun-like star with with a tidally-locked extrasolar planet 
(e.g., $\tau$ Boo; \citep{saar04,catala07}) located at the same distance as before. 

In both simulations, the boundary condition for the planetary plasma number density and temperature were 
$n_0=10^{10}~cm^{-3}$ and $T_0=10^4~K$ respectively \citep{murray-clay09}. The stellar boundary conditions were 
$n_0=10^9~cm^{-3}$ and $T_0=3.5\cdot10^6~K$ respectively, based on previous simulation of the solar corona 
\citep{cohen07}. To further aide interpretation of the results, we performed two additional simulations 
as a reference, identical to the cases above, but with the planet removed; i.e., considering the star 
with just the $5G$ dipolar field, and with the CR2010 magnetogram.

In each simulation, the set of MHD equations is solved until convergence. The end result is a 
three-dimensional, steady-state solution for the particular system that includes all the MHD variables 
(density, pressure, velocity, and magnetic field). Since the MHD solution contains the values for $n_e$ and 
$T$ at each spatial cell, we can perform the line-of-sight integration to obtain the predicted X-ray emissions 
for a particular view angle. The integration takes in to account cells in front of the star but omits cells 
behind it. We repeat this procedure for different view angles to mimic the predicted X-ray flux as the system 
rotates. The X-ray flux, $f_x$, is calculated as the line-of-sight 
integral $f_x\approx \int n_e^2\cdot P(T)dl$. Here $dl$ is the line-of-sight depth and we have used a piecewise
linear approximation to the radiative loss $P(T)$ for a plasma with solar photospheric abundances 
\citep{giampapa96}. 

%%%%%%%%%%%%%%%%%%%%%%%%%%%%%%%%%%%%%%%%%%%%%%%%%%%%%%%%%%%%%%%%%%%%%%%%%%%%%%
% Section - Results
%%%%%%%%%%%%%%%%%%%%%%%%%%%%%%%%%%%%%%%%%%%%%%%%%%%%%%%%%%%%%%%%%%%%%%%%%%%%%%

\section{RESULTS}
\label{sec:Results}

\begin{figure*}[h!]
\centering
\includegraphics[width=3.0in]{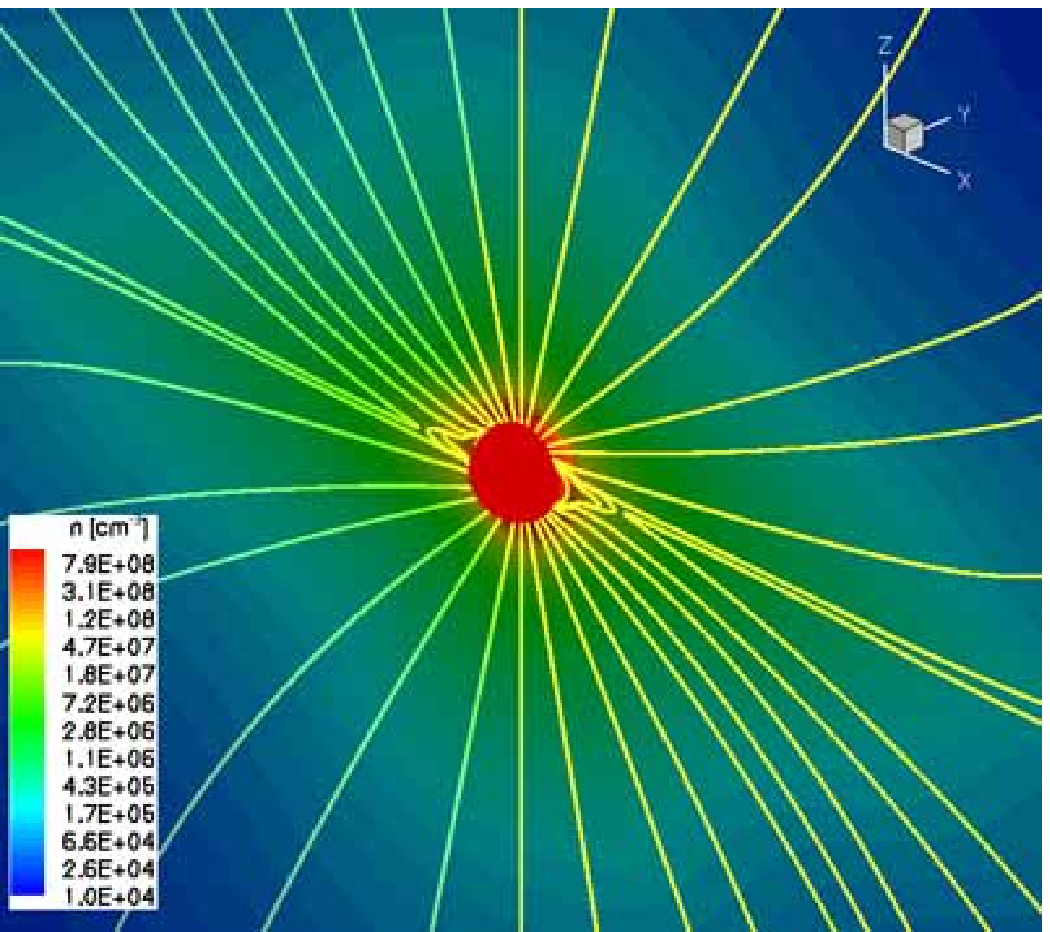}
\includegraphics[width=3.0in]{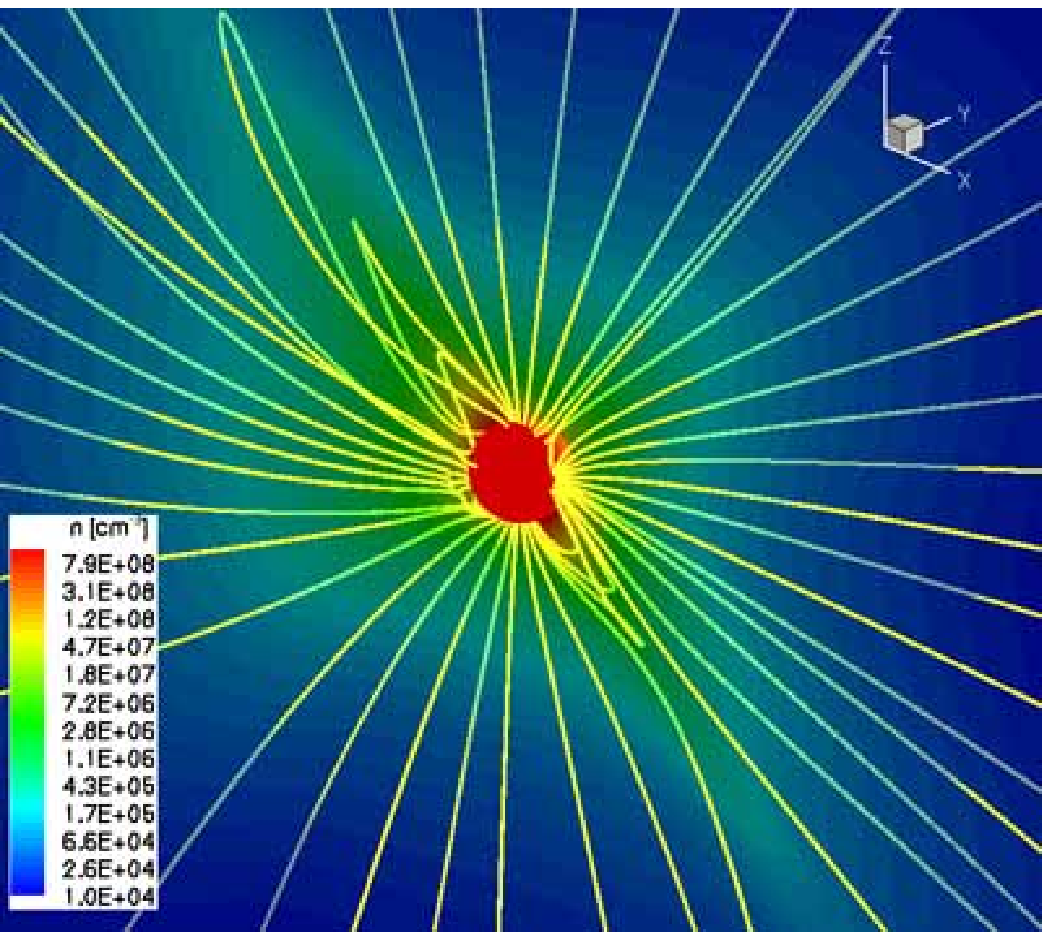} \\
\includegraphics[width=3.0in]{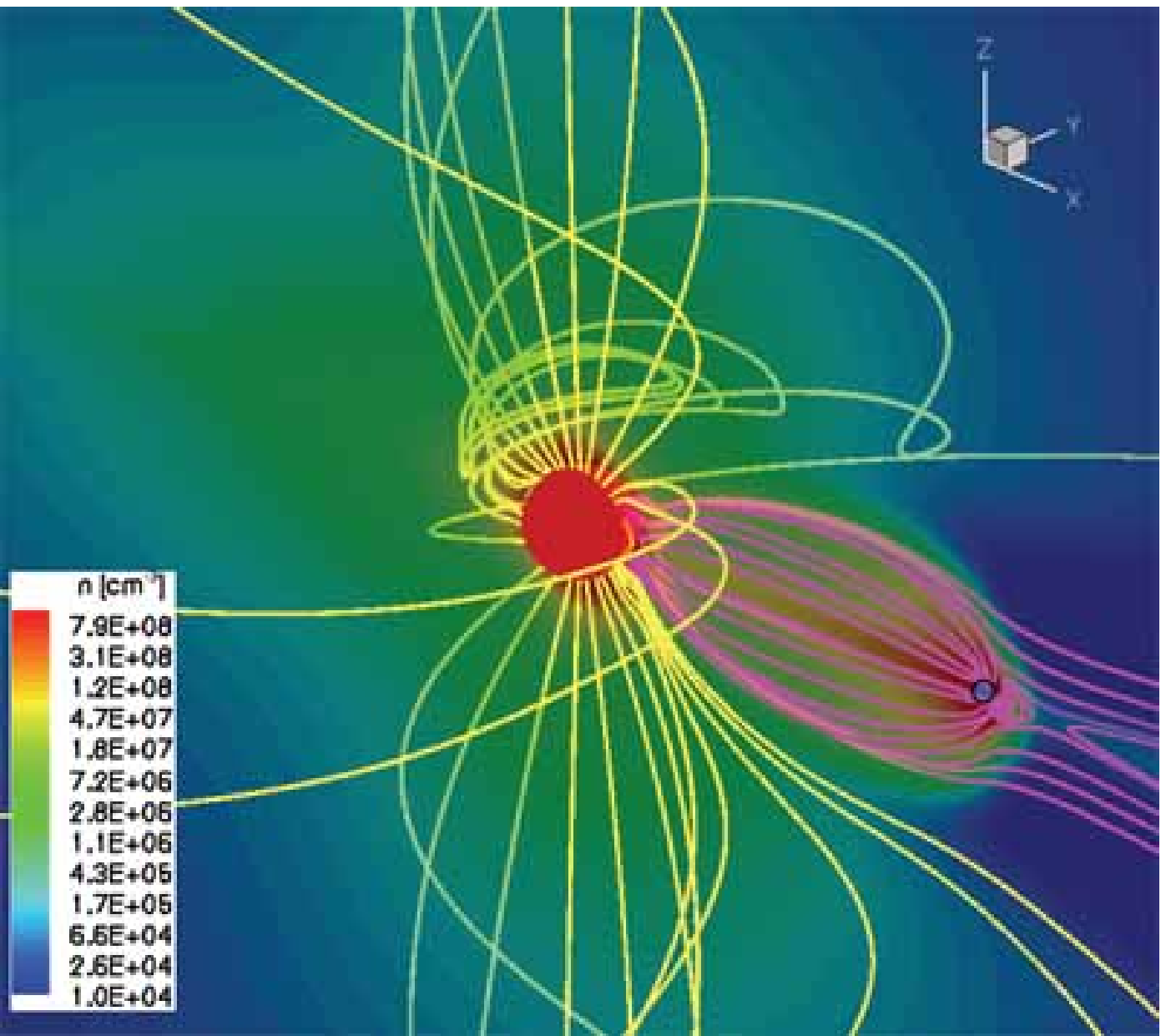}
\includegraphics[width=3.0in]{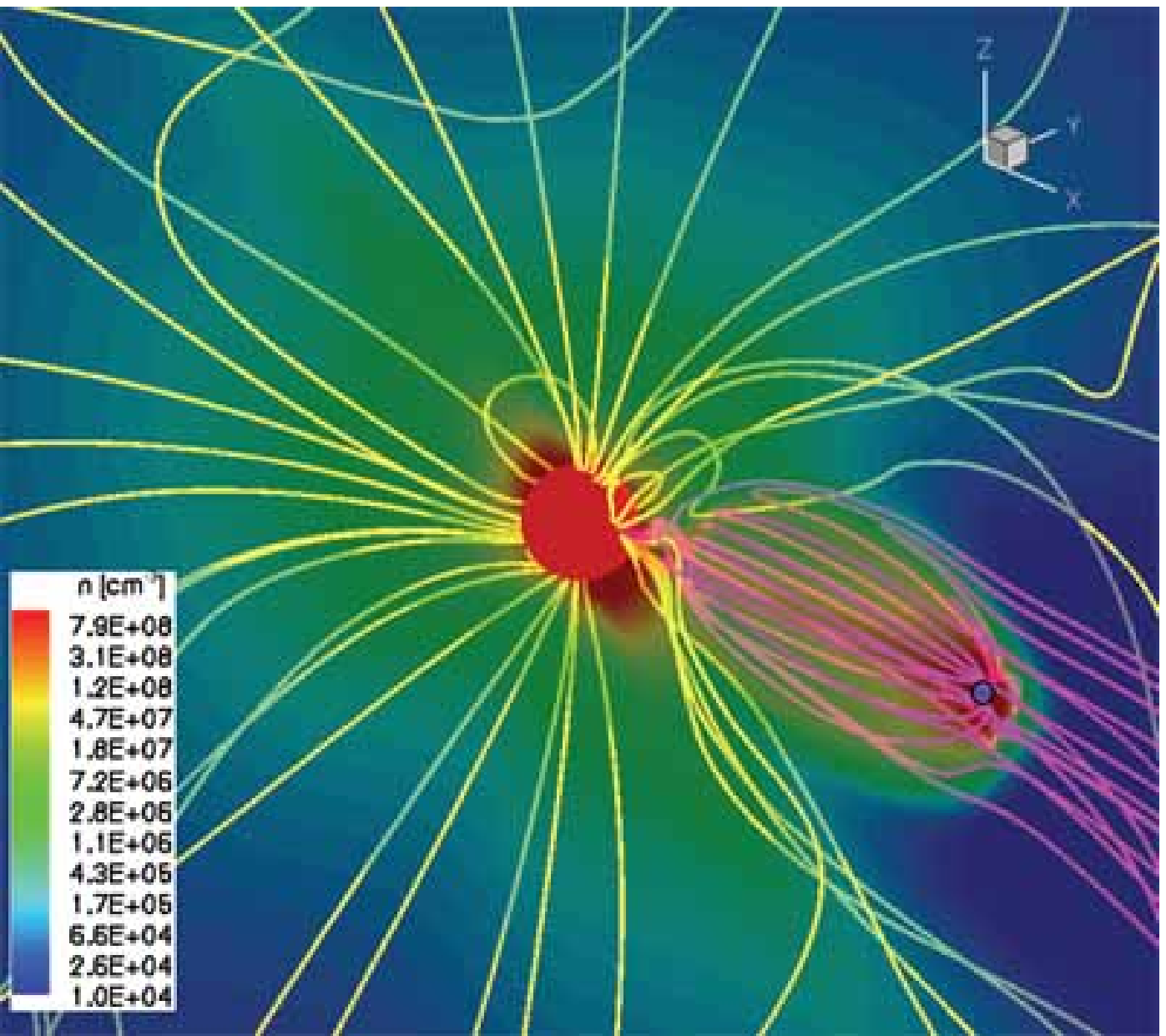} \\
\includegraphics[width=3.0in]{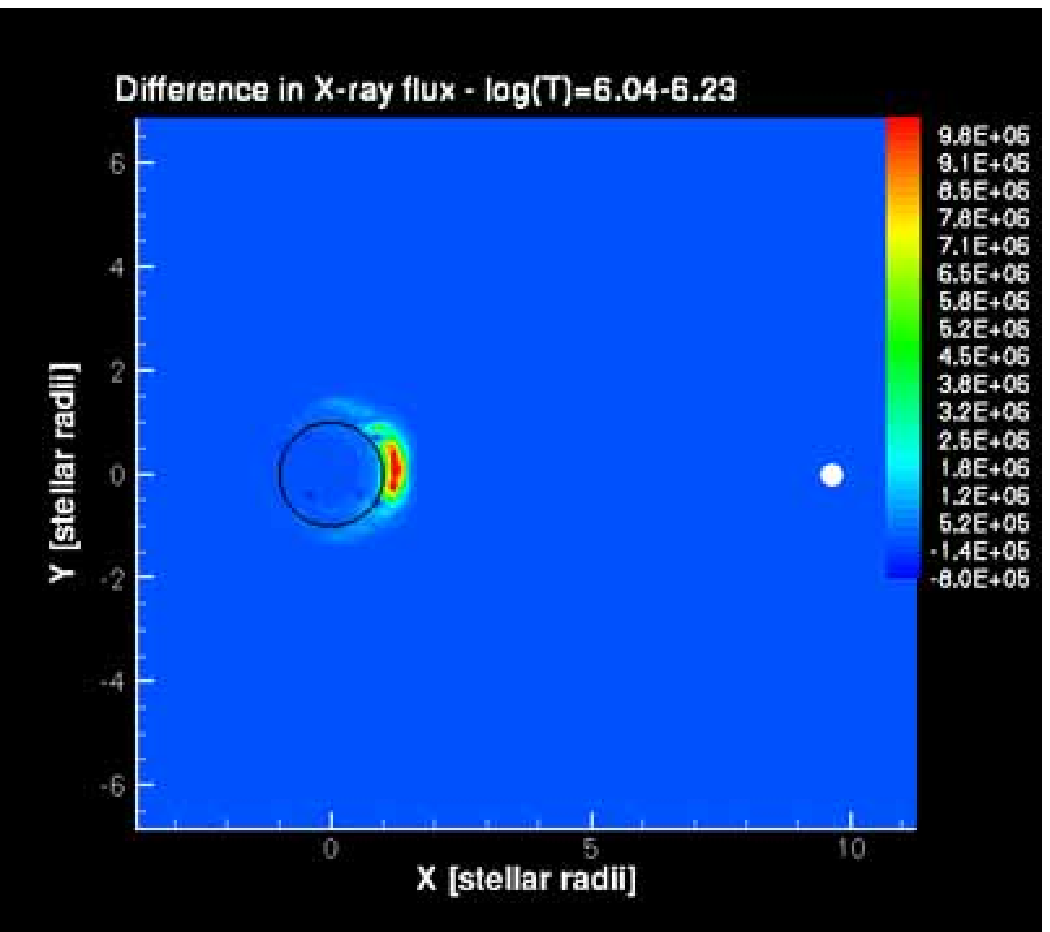}
\includegraphics[width=3.0in]{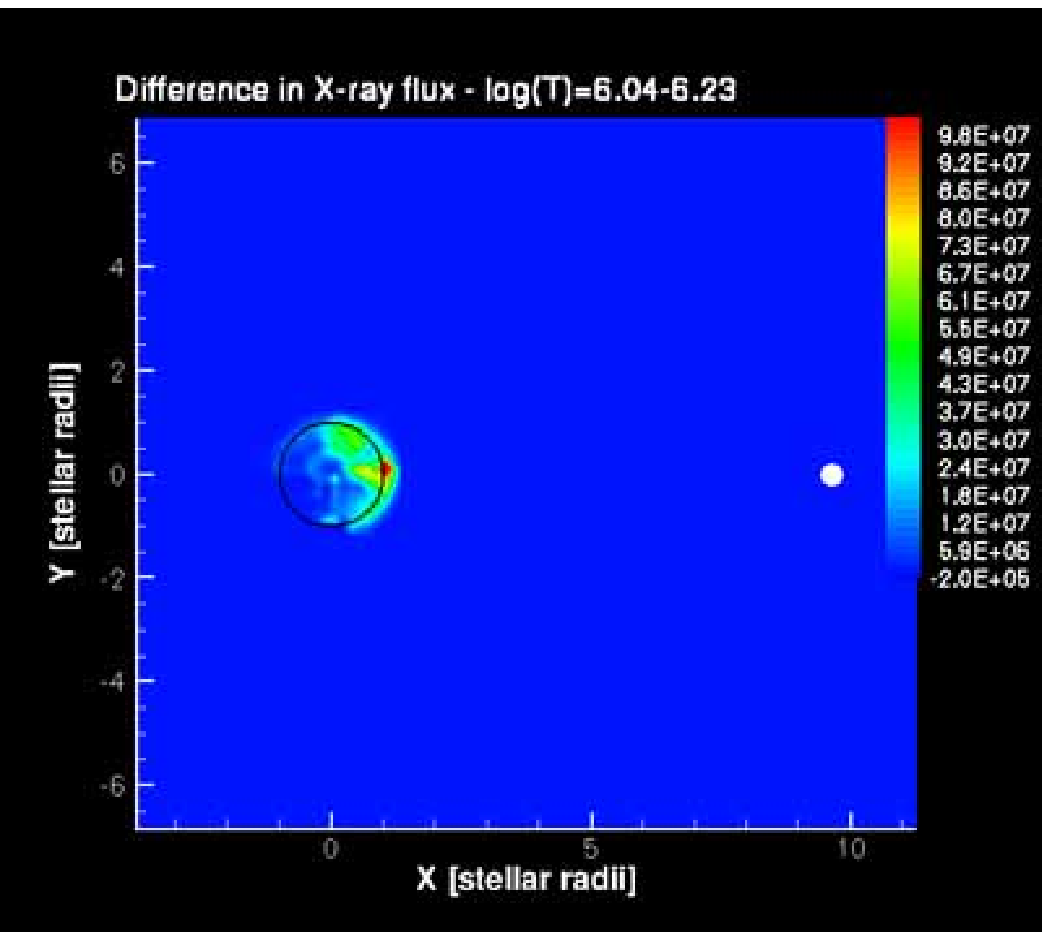} 
\caption{Illustration of the MHD simulation results for Case A (left), 
and for Case B (right). The top panels show the reference simulations without the planet, while the 
simulations including the planet are shown in the middle panels (planetary surface is marked by black circle). 
Color contours represent plasma number density on the y=0 plane, and yellow three-dimensional streamlines 
represent stellar magnetic 
field lines that originate on the intersection of the y=0 plane with the stellar surface. The inner 
boundary of the simulation domain (the stellar surface) is represented by a red sphere, and the view 
is from an angle of $45^\circ$. In the middle panels, the magnetic field lines that connect to the planetary 
magnetosphere are drawn in pink. The bottom panels show the difference in X-ray emission for plasma 
in the logarithmic temperature range $\log{T}=6.04-6.23$ between the case with and without the planet. 
Black circles represent the stellar surface and indicate that the 'hot spots' are located in the low 
corona.}
\label{fig:f1}
\end{figure*}

Simulation results are illustrated in Figure~\ref{fig:f1}. Top and middle panels show the three-dimensional solutions 
excluding and including the planet, respectively. Bottom panels show the difference in X-ray flux for temperature range 
of $log(T)=6.04-6.23$ with and without the planet. Left panels show results of Case A, while right panels show results 
of Case B. 

Considering first Case A, we note a key 
difference between simulations with and without the planet. In the former, magnetic field lines are conspicuously 
brought in toward the planet and are constricted due to the presence of the planetary magnetosphere. This has a 
palpable influence on the coronal electron number density, $n_e$, which now increases azimuthally approaching the 
star-planet line to form an X-ray 'bright spot' facing the planet. This solution is qualitatively very similar 
in Case B, where there is also a clear longitudinal concentration of the plasma density, and consequently the 
X-ray flux (which is proportional to $n^2_e$). 
While we cannot simulate chromospheric emission with our current models, its surface intensity distribution 
on the Sun follows closely those regions of the disk that are brighter at EUV and X-ray wavelengths. The 
results of these simulations are, then, fully consistent with the observed location of chromospheric hot 
spots seen in-phase with the planetary orbit.

\begin{figure*}[h!]
\centering
\includegraphics[width=7in]{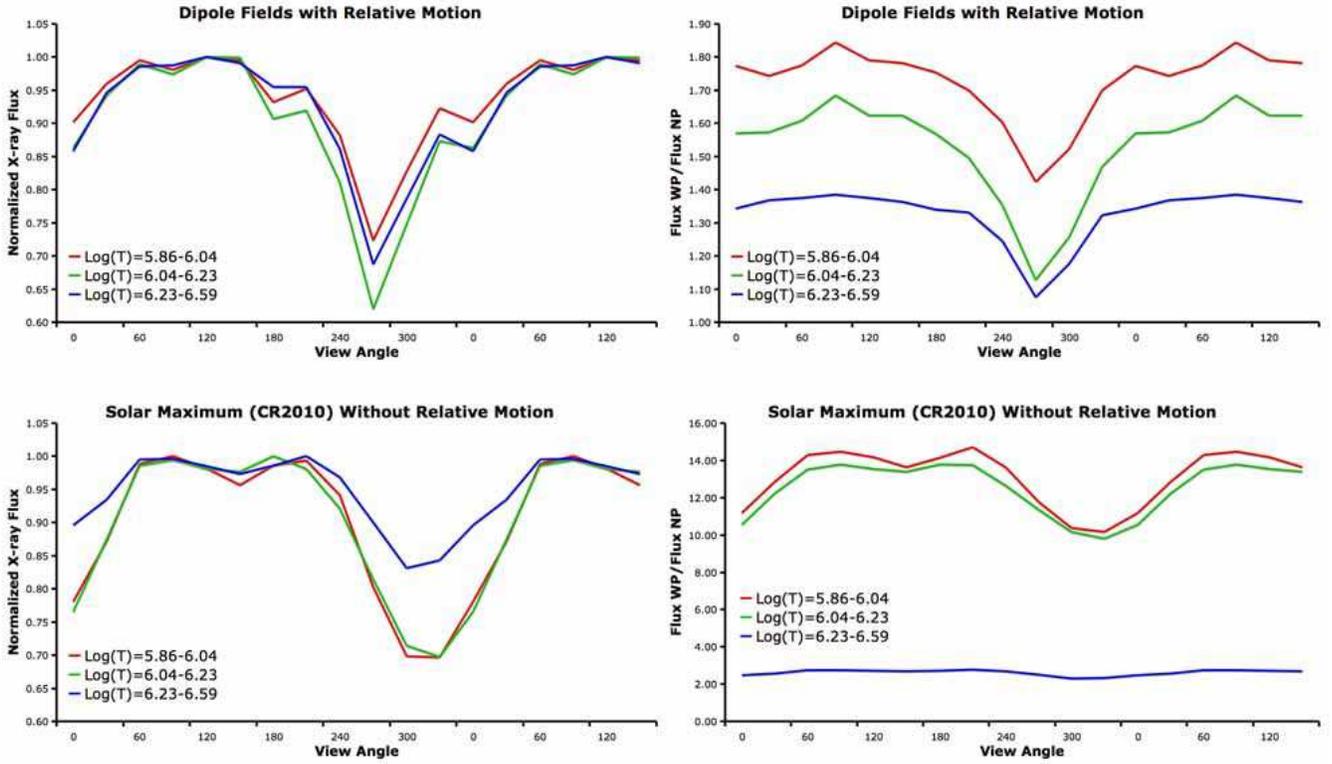}
\caption{Integrated LOS X-ray flux for the dipolar case with relative orbital motion (top) and for 
the realistic solar maximum case (bottom) as a function of view angle. The planet is located at $90^\circ$ 
and is hidden by the star at $270^\circ$. Left panels show the flux for three temperature bins, where each 
curve is normalized to its own maximum value. Right panels show the ratio of the X-ray flux with and 
without the planet for each temperature bin as a function of view angle.}
\label{fig:f2}
\end{figure*}

The longitudinal brightening effect is also clear in Figure~\ref{fig:f2}, where the line-of-sight X-ray flux originating at 
different plasma temperatures is shown as a function of viewing angle for the simulations including planets. In 
the idealized dipole field case, the emission is much more intense when the corona is viewed from the direction 
of the planet than from the opposite direction when the star hides the brighter parts of the corona that form the 
hot spot in phase with the planet (a difference of 25-35\% in observed X-ray flux). The simulation based on the realistic, 
complex magnetic field results in a hot spot (15-30\% difference in X-ray flux) shifted by about $60^\circ$ relative to 
the star-planet line. This suggests that the phase shifts between hot spots and planetary orbital phase seen 
in SPI observations are probably due to the complexity of the stellar coronal magnetic fields and the consequent 
complexity of the magnetic connectivity between star and the planet.

Also of interest is the magnitude of the X-ray flux enhancement seen compared to the case where there is no planet. 
In the ideal dipole field case, we find enhancements of at least 10\% in base emission and as much as $\approx$80\% 
relative to the no-planet case, and a contrast between 
minimum and maximum of ~20-30\%. These numbers are consistent with the observed signatures of SPI found in X-ray 
observations. When solutions driven by realistic magnetogram data are considered, we find that enhancements in 
base emission by a factor of 10 are possible. The simulation with realistic, complex magnetic field results in much 
higher density enhancement and closing of coronal loops compared to the dipolar case.

The MHD model, however, does not take into account the detailed physics of smaller 
scale X-ray emitting coronal loops, and the magnitude of the predicted X-ray enhancements should be considered 
approximate. The density enhancement in the closed-field zone provides a vital medium which can be heated to produce the SPI 
effect. The compression alone produces significant increase in radiative loss at low temperatures; localized dynamical 
effects, not modeled here, such as megnetic reconnection due to the planetary orbital motion, and particle acceleration 
then likely further heat the plasma to the observed $\sim$ 1 keV temperatures \citep{saar08}.

%%%%%%%%%%%%%%%%%%%%%%%%%%%%%%%%%%%%%%%%%%%%%%%%%%%%%%%%%%%%%%%%%%%%%%%%%%%%%%
% Section - Conclusions
%%%%%%%%%%%%%%%%%%%%%%%%%%%%%%%%%%%%%%%%%%%%%%%%%%%%%%%%%%%%%%%%%%%%%%%%%%%%%%

\section{Conclusions}
\label{sec:Conclusions}
In conclusion, we find that a dominant physical effect creating observable time-variable spi signatures is that 
the existence of the planet and its magnetosphere, 
close to the star, prevents the expansion of the stellar coronal magnetic field and the acceleration of the stellar 
wind. The pressure gradient is not as large as it would be in the absence of the planet, so the coronal field 
lines that would be opened by the wind remain closed and the plasma in these loops do not escape. This effect alone 
reproduces three observable feature: 1) Enhancement of total X-ray flux, 2) Appearance of coronal hot spots, 3) Phase-shift 
of the hot spots from star-planet line. The density enhancement results in low temperature coronal heating. We will 
further develop the model to include the planetary orbital motion in order to capture more dynamical effects of SPI.

%%%%%%%%%%%%%%%%%%%%%%%%%%%%%%%%%%%%%%%%%%%%%%%%%%%%%%%%%%%%%%%%%%%%%%%%%%%%%%
% Acknowledgments
%%%%%%%%%%%%%%%%%%%%%%%%%%%%%%%%%%%%%%%%%%%%%%%%%%%%%%%%%%%%%%%%%%%%%%%%%%%%%%

\acknowledgments
This work has been inspired by initial study performed by Noe Lugaz. We thank 
an unknown referee for his/hers useful comments and Ruth Murray-Clay for useful discussion. 
OC is supported by NSF-SHINE ATM-0823592 grant, NASA-LWSTRT Grant NNG05GM44G. 
JJD and VLK were funded by NASA contract NAS8-39073 to 
the {\it Chandra X-ray Center}.  Simulation results were obtained using the Space Weather 
Modelling Framework, developed by the Canter for Space Environment Modelling, at the 
University of Michigan with funding support from NASA ESS, NASA ESTO-CT, NSF KDI, 
and DoD MURI.

%\bibliographystyle{apj}
%\bibliography{ESP.bib}

\begin{thebibliography}{23}
\expandafter\ifx\csname natexlab\endcsname\relax\def\natexlab#1{#1}\fi

\bibitem[{{Catala} {et~al.}(2007){Catala}, {Donati}, {Shkolnik}, {Bohlender},
  \& {Alecian}}]{catala07}
{Catala}, C., {Donati}, J.-F., {Shkolnik}, E., {Bohlender}, D., \& {Alecian},
  E. 2007, \mnras, 374, L42

\bibitem[{{Cohen} {et~al.}(2007){Cohen}, {Sokolov}, {Roussev}, {Arge},
  {Manchester}, {Gombosi}, {Frazin}, {Park}, {Butala}, {Kamalabadi}, \&
  {Velli}}]{cohen07}
{Cohen}, O., {Sokolov}, I.~V., {Roussev}, I.~I., {Arge}, C.~N., {Manchester},
  W.~B., {Gombosi}, T.~I., {Frazin}, R.~A., {Park}, H., {Butala}, M.~D.,
  {Kamalabadi}, F., \& {Velli}, M. 2007, Astrophys. J., 654, L163

\bibitem[{{Cohen} {et~al.}(2008){Cohen}, {Sokolov}, {Roussev}, \&
  {Gombosi}}]{cohen08}
{Cohen}, O., {Sokolov}, I.~V., {Roussev}, I.~I., \& {Gombosi}, T.~I. 2008,
  Journal of Geophysical Research (Space Physics), 113, 3104

\bibitem[{{Cranmer} \& {Saar}(2007)}]{cranmer07}
{Cranmer}, S.~R., \& {Saar}, S.~H. 2007, ArXiv Astrophysics e-prints

\bibitem[{{Cuntz} {et~al.}(2000){Cuntz}, {Saar}, \& {Musielak}}]{cuntz00}
{Cuntz}, M., {Saar}, S.~H., \& {Musielak}, Z.~E. 2000, Astrophys. J., 533, L151

\bibitem[{{Giampapa} {et~al.}(1996){Giampapa}, {Rosner}, {Kashyap}, {Fleming},
  {Schmitt}, \& {Bookbinder}}]{giampapa96}
{Giampapa}, M.~S., {Rosner}, R., {Kashyap}, V., {Fleming}, T.~A., {Schmitt},
  J.~H.~M.~M., \& {Bookbinder}, J.~A. 1996, Astrophys. J., 463, 707

\bibitem[{{Grie{\ss}meier} {et~al.}(2004){Grie{\ss}meier}, {Stadelmann},
  {Penz}, {Lammer}, {Selsis}, {Ribas}, {Guinan}, {Motschmann}, {Biernat}, \&
  {Weiss}}]{griessmeier04}
{Grie{\ss}meier}, J.-M., {Stadelmann}, A., {Penz}, T., {Lammer}, H., {Selsis},
  F., {Ribas}, I., {Guinan}, E.~F., {Motschmann}, U., {Biernat}, H.~K., \&
  {Weiss}, W.~W. 2004, Astron. \& Astrophys., 425, 753

\bibitem[{{Kashyap} {et~al.}(2008){Kashyap}, {Drake}, \& {Saar}}]{kashyap08}
{Kashyap}, V.~L., {Drake}, J.~J., \& {Saar}, S.~H. 2008, Astrophys. J., 687,
  1339

\bibitem[{{Lanza}(2008)}]{lanza08}
{Lanza}, A.~F. 2008, Astron. \& Astrophys., 487, 1163

\bibitem[{{Lanza}(2009)}]{lanza09}
---. 2009, ArXiv e-prints

\bibitem[{{Mayor} {et~al.}(2003){Mayor}, {Naef}, {Pepe}, {Queloz}, {Santos}, \&
  {Udry}}]{exoplanets03}
{Mayor}, M., {Naef}, D., {Pepe}, F., {Queloz}, D., {Santos}, N., \& {Udry}, S.
  2003, {The Geneva extrasolar planet search programmes},
  {http://exoplanets.eu}

\bibitem[{{Mayor} \& {Queloz}(1995)}]{mayor95}
{Mayor}, M., \& {Queloz}, D. 1995, Nature, 378, 355

\bibitem[{{Murray-Clay} {et~al.}(2009){Murray-Clay}, {Chiang}, \&
  {Murray}}]{murray-clay09}
{Murray-Clay}, R.~A., {Chiang}, E.~I., \& {Murray}, N. 2009, Astrophys. J.,
  693, 23

\bibitem[{{Olson} \& {Christensen}(2006)}]{olson06}
{Olson}, P., \& {Christensen}, U.~R. 2006, Earth and Planetary Science Letters,
  250, 561

\bibitem[{{Powell} {et~al.}(1999){Powell}, {Roe}, {Linde}, {Gombosi}, \& {de
  Zeeuw}}]{powell99}
{Powell}, K.~G., {Roe}, P.~L., {Linde}, T.~J., {Gombosi}, T.~I., \& {de Zeeuw},
  D.~L. 1999, Journal of Computational Physics, 154, 284

\bibitem[{{Saar} {et~al.}(2008){Saar}, {Cuntz}, {Kashyap}, \& {Hall}}]{saar08}
{Saar}, S.~H., {Cuntz}, M., {Kashyap}, V.~L., \& {Hall}, J.~C. 2008, in IAU
  Symposium, Vol. 249, IAU Symposium, ed. Y.-S. {Sun}, S.~{Ferraz-Mello}, \&
  J.-L. {Zhou}, 79--81

\bibitem[{{Saar} {et~al.}(2004){Saar}, {Cuntz}, \& {Shkolnik}}]{saar04}
{Saar}, S.~H., {Cuntz}, M., \& {Shkolnik}, E. 2004, in IAU Symposium, Vol. 219,
  Stars as Suns : Activity, Evolution and Planets, ed. A.~K. {Dupree} \& A.~O.
  {Benz}, 355--+

\bibitem[{{S{\'a}nchez-Lavega}(2004)}]{sanches-lavega04}
{S{\'a}nchez-Lavega}, A. 2004, Astrophys. J., 609, L87

\bibitem[{{Shkolnik} {et~al.}(2008){Shkolnik}, {Bohlender}, {Walker}, \&
  {Collier Cameron}}]{shkolnik08}
{Shkolnik}, E., {Bohlender}, D.~A., {Walker}, G.~A.~H., \& {Collier Cameron},
  A. 2008, Astrophys. J., 676, 628

\bibitem[{{Shkolnik} {et~al.}(2003){Shkolnik}, {Walker}, \&
  {Bohlender}}]{shkolnik03}
{Shkolnik}, E., {Walker}, G.~A.~H., \& {Bohlender}, D.~A. 2003, \apj, 597, 1092

\bibitem[{{Shkolnik} {et~al.}(2005{\natexlab{a}}){Shkolnik}, {Walker},
  {Bohlender}, {Gu}, \& {Kurster}}]{shkolnik05a}
{Shkolnik}, E., {Walker}, G.~A.~H., {Bohlender}, D.~A., {Gu}, P.-G., \&
  {Kurster}, M. 2005{\natexlab{a}}, Astrophys. J., 622, 1075

\bibitem[{{Shkolnik} {et~al.}(2005{\natexlab{b}}){Shkolnik}, {Walker},
  {Rucinski}, {Bohlender}, \& {Davidge}}]{shkolnik05b}
{Shkolnik}, E., {Walker}, G.~A.~H., {Rucinski}, S.~M., {Bohlender}, D.~A., \&
  {Davidge}, T.~J. 2005{\natexlab{b}}, Astronom. J., 130, 799

\bibitem[{{T{\'o}th} {et~al.}(2005){T{\'o}th}, {Sokolov}, {Gombosi}, {Chesney},
  {Clauer}, {De Zeeuw}, {Hansen}, {Kane}, {Manchester}, {Oehmke}, {Powell},
  {Ridley}, {Roussev}, {Stout}, {Volberg}, {Wolf}, {Sazykin}, {Chan}, {Yu}, \&
  {K{\'o}ta}}]{toth05}
{T{\'o}th}, G., {Sokolov}, I.~V., {Gombosi}, T.~I., {Chesney}, D.~R., {Clauer},
  C.~R., {De Zeeuw}, D.~L., {Hansen}, K.~C., {Kane}, K.~J., {Manchester},
  W.~B., {Oehmke}, R.~C., {Powell}, K.~G., {Ridley}, A.~J., {Roussev}, I.~I.,
  {Stout}, Q.~F., {Volberg}, O., {Wolf}, R.~A., {Sazykin}, S., {Chan}, A.,
  {Yu}, B., \& {K{\'o}ta}, J. 2005, Journal of Geophysical Research (Space
  Physics), 110, 12226

\end{thebibliography}

\clearpage

\end{document}